\documentclass[twocolumn,epsfig,floats,aps]{revtex4}
\usepackage{amsmath,bm}
\usepackage{epsfig}
\usepackage{graphicx}

\begin{document}
\title{Reexamination for the calculation of elliptic flow and 
       other fourier harmonics}
\author{Xiao-Mei Li$^1$, Bao-Guo Dong$^1$, Yu-Liang Yan$^1$, Hai-Liang Ma$^1$, 
Dai-Mei Zhou$^2$, and Ben-Hao Sa$^{1,2,3}$ \footnote{Corresponding author: 
sabh@ciae.ac.cn}}
\affiliation{
$^1$  China Institute of Atomic Energy, P. O. Box 275 (18),
      Beijing, 102413 China \\
$^2$  Institute of Particle Physics, Huazhong Normal University,
      Wuhan, 430079 China \\
$^3$  CCAST (World Lab.), P. O. Box 8730 Beijing, 100080 China}
\begin{abstract}
We have argued that the azimuthal symmetry and asymmetry components in fourier 
expansion of particle momentum azimuthal distribution, $v_n$ (n=0, 1, 2, ...), 
should be calculated as an average of $cos(n\phi)$ first over particles in an 
event and then over events (event-wise average) rather than ``an average over 
all particles in all events" (particle-wise average). In case of large 
centrality (multiplicity) bin the particle-wise average is not accurate 
because the influence (fluctuation) of particle multiplicity was not taken
into account.
\\
\noindent{PACS numbers: 25.75.Dw, 24.85.+p}
\end{abstract}
\maketitle
Elliptic flow (proportional to $v_2$) and other harmonics (proportional to 
$v_n$, $n$=0, 1, 3, 4, ...) as fourier expansion coefficients of azimuthal 
distribution of particle momentum are highly sensitive to the eccentricity of 
the early created fireball in ultra-relativistic heavy ion collisions. The 
expected phase transition to Quark-Gluon-Plasma (QGP) should have a dramatic 
effect on those harmonics. The consistency between experimental data of $v_2
(p_T)$ at mid-rapidity and the corresponding hydrodynamic predictions are 
regarded as an evidence of the production of partonic matter in 
ultra-relativistic nucleus-nucleus collisions \cite{miclo}. The elliptic flow 
of high $p_T$ particles may be related to the jet fragmentation and parton 
energy loss \cite{wang1}, which are usually not included in the hydrodynamic 
calculations. This hydrodynamic model \cite{kolb} overestimates $v_2(p_T)$ in 
the $p_T \geq 1.5$ GeV/c region \cite{phob1}. That is regarded, together with 
the discovery of jet quenching \cite{zajc}, as a strong evidence of sQGP 
formation in relativistic nucleus-nucleus collisions at RHIC.

The elliptic flow and other harmonics have become an interesting topic in the 
field of ultra-relativistic nucleus-nucleus collisions. A lot of experimental 
data from RHIC have been published \cite{star1,phen1,phob2}. Consequently the
microscopic transport model studies are also widely progressing 
\cite{bin1,fuchs,chen1,xu1,zhu} as well as the abundant hydrodynamic 
investigations.

Refs. \cite{zhang,posk} are two well known pioneering papers in this field.
In \cite{posk} the investigation was starting from the triple differential
distribution of
\begin{equation}
E{\frac{d^3N}{d^3p}}=\frac{1}{2\pi}{\frac{d^2N}{p_Tdydp_T}}
       [1+\sum_{n=1,...}2v_ncos[n(\phi-\Psi_r)]] \label{eq14}.
\end{equation}
Then it was defined that: ``... $v_n=\langle cos[n(\phi-\Psi_r)]\rangle$, 
where $\langle \rangle$ indicates an average over all particles in all events. 
For the particle number distribution, the coefficient $v_1$ is $\langle p_x/
p_t\rangle$ and $v_2$ is $\langle(p_x/p_t)^2-(p_y/p_t)^2\rangle$." (This kind 
of average will be indicated as particle-wise average hereafter.) Later that 
is widely accepted in theoretical calculations and becomes a common 
convention: the zeroth harmonic is $v_0 =1$ and $n$-$th$ harmonic is $v_n=
\langle cos[n(\phi-\Psi_r)]\rangle$ ($n$=1, 2, 3, ... ), here $\langle\rangle$ 
indicates particle-wise average.  

As mentioned above that the elliptic flow and other harmonics are important 
and widely studied for over a decade. But we have to review the calculation of 
$v_n$ in this paper: does the $v_n$ should be calculated as an average of 
$cos(n\phi)$ first over particles in an event and then over events (event-wise 
average) or calculated as an average of $cos(n\phi)$ over all particles in all 
events (particle-wise average). To this end, we first rederive the elliptic 
flow and other harmonics starting from the particle number (multiplicity) 
distribution. For that deduction one has first to define the reaction plane.
 
\begin{figure}[h]
\centerline{\hspace{-0.5in}
\epsfig{file=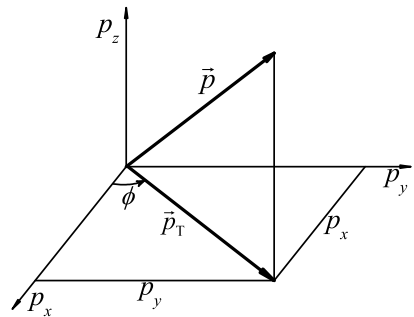,height=1.8in,angle=0}}
%\vspace{0.2in}
\vspace{-0.1in} \caption{The particle momentum $\vec p$, in the
Cartesian coordinate system,[$p_x$,$p_y$,$p_z$].} \label{pphi}
\end{figure}

In theory, if the beam direction and impact parameter vector are fixed, 
respectively, at the $p_z$ and $p_x$ axes, then the reaction plane is just the 
$p_x-p_z$ plane \cite{zhang}. Therefore the reaction plane angle ($\Psi_r$) 
between reaction plane and the $p_x$ axis \cite{zhang} introduced for the 
extraction of elliptic flow in the experiments \cite{posk} is zero. The 
particle azimuthal angle ($\phi$) in momentum space is measured with respect 
to the reaction plane, which is consistent with the definition in \cite{posk}. 
This particle azimuthal angle ($\phi$) is just the angle spanned by $\vec p_T$ 
relative to the $p_x$ axis as shown in Fig. \ref{pphi}.

However, in experiment, the reaction plane is different event by event, thus 
the experimental measurement of elliptic flow is not trivial. One has to 
invoke a complex reaction plane identification method (two-particle 
correlation method) \cite{posk}, cumulant method \cite{borg}, or Lee-Yang 
zeroes method \cite{lee}. In all of these methods a quantity has to be first 
constructed event by event: the event plane in \cite{posk}, cumulant expansion 
of weighted $n$-$th$ transverse event-flow vector in \cite{borg}, and a 
generating function in \cite{lee}. Then a corresponding average over measured 
events has to be taken. Therefore the experimental extraction of elliptic flow 
parameter is event-wise average indeed.   

The particle number (multiplicity) distribution can be expressed as
\begin{equation}
E{\frac{d^3N}{d^3p}}={\frac{d^3N}{p_Tdydp_Td\phi}},
\end{equation}
in the cylindrical coordinate system after substituting $p_z$ by y (rapidity) 
and using the relation of $dy/dp_z=1/E$ \cite{pdg}. Then the normalized 
particle multiplicity distribution is
\begin{equation*}
{\frac{1}{N_{ev}}}{\frac{d^3N}{p_Tdydp_Td\phi}},
\end{equation*}
\begin{equation}
N_{ev}=\int dy\int p_Tdp_T\int d\phi{\frac{d^3N}{p_Tdydp_Td\phi}},
\label{eq11}
\end{equation}
where $N_{ev}$ is the event multiplicity and the integrals are taken over 
entire range of the variables.

This normalized particle multiplicity distribution in Eq.(\ref{eq11}) is a 
three dimensional ($y, p_T, \phi$) distribution function. The dimension can be 
reduced by integrating over a certain variable \cite{reic}. For example, to 
study the $v_n$ as a function of rapidity $y$, $v_n(y)$, one should take 
integral over $p_T$, then above three dimensional distribution function 
reduces to a two dimensional distribution function
\begin{equation}
{\frac{1}{N_{ev}}}{\int p_Tdp_T \frac{d^3N}{p_Tdydp_Td\phi}}.
\end{equation}
In numerical calculations a small rapidity interval, $\Delta y$, is used 
instead of single $y$ value, the corresponding normalized particle 
multiplicity distribution becomes
\begin{equation*}
{\frac{1}{N_{evp}}}{\int_pdy'\int p_Tdp_T \frac{d^3N}{p_Tdy'dp_Td\phi}}
\end{equation*}
\begin{equation*}
=\frac{d}{d\phi}[{\frac{1}{N_{evp}}}{\int_pdy'\int p_Tdp_T \frac{d^2N}{p_Tdy'dp_T}}]
\end{equation*}
\begin{equation*}
\equiv f(\phi),
\end{equation*}
where
\begin{equation*}
 N_{evp}=\int_pdy'\int p_Tdp_T\int d\phi{\frac{d^3N}{p_Tdy'dp_Td\phi}},
\end{equation*}
and
%\begin{center}
\begin{equation}
\int_p dy' \equiv \int_{y-\Delta y/2}^{y+\Delta y/2}dy'.
\label{eq6}
\end{equation}
%\end{center}
%\begin{widetext}
%\begin{center}
Note that, the normalization factor $N_{evp}$ here is different from $N_{ev}$ 
in Eq. (\ref{eq11}) in the range of integral over $y$ and is denoted as 
constrained event multiplicity. In the above equation $f(\phi)$ is the 
normalized distribution density function of $\phi$ and the $N_{evp}f(\phi)$ is 
the number of particles emitted into $d\phi$ at azimuthal angle $\phi$ without 
constraint on $p_T$ but $y$ is constrained in $\Delta y$. This can be 
constructed experimentally and/or calculated theoretically. Here we have taken 
$v_n(y)$ as an example but it is similar for the $v_n(p_T)$.

Since $f(\phi)$ is $2\pi$ periodic and even function of $\phi$, it can be 
expanded by a fourier series \cite{beyer} as
\begin{equation}
%f(\phi)=\frac{1}{2\pi}[1+\sum_{n=1}2v_ncos(n\phi)],
f(\phi)=\frac{1}{\pi}[\frac{1}{2}+\sum_{n=1}v_ncos(n\phi)],
\label{eq4}
\end{equation}
%\begin{widetext}
\vspace{0.005cm}
\begin{equation*}
v_n=\int d\phi f(\phi)cos(n\phi)
\end{equation*}
\begin{equation}
\hspace{-3.9cm} \hspace{5cm} \equiv \overline{cos(n\phi)}
{\hspace{0.2cm} (n=1, 2, ...).}
\label{eq7}
\end{equation}
%\end{widetext}
In above equation, $\overline{cos(n\phi)}$ denotes the average of $cos(n\phi)$ 
over particles in a single event and can be calculated both experimentally and 
theoretically. The first expansion term, in Eq.(\ref{eq4}), is a circle 
(isotropic), second term ($cos(\phi)$) a leave (directed flow), third term 
($cos(2\phi)$) a four-leaved rose (elliptic flow), and the $(n+1)$-$th$ term 
($cos(n\phi)$) a multi-leaved rose ($n$-$th$ harmonic) \cite{beyer}.

As the event multiplicity fluctuates event-by-event, one always has to 
generate multiple events and to take average over events generated. So the 
$v_n$ in Eq.(\ref{eq7}) should be
\begin{equation}
v_n^e=\langle\overline{cos(n\phi)}\rangle_{ev}{\hspace{0.2cm} (n=1, 2, ...),}
\label{eq3}
\end{equation}
where $\langle...\rangle_{ev}$ means an average over all events. A superscript
, ``$e$", is here added on $v_n$ to identify this average as the event-wise 
average. Applying recursion formula of cosine function \cite{beyer}
\begin{equation}
cos(n\phi)=2cos((n-1)\phi)cos \phi-cos((n-2)\phi)
\end{equation}
and
\begin{equation}
cos \phi=\frac{p_x}{p_T},  \hspace{0.2cm} sin \phi=\frac{p_y}{p_T},
\end{equation}
to Eq. (\ref{eq3}), $v_n^e$ can be expressed as
\begin{eqnarray}
v_1^e=\langle\overline{[\frac{p_x}{p_T}]}\rangle_{ev},\hspace{0.2cm}
v_2^e=\langle\overline{[\frac{p_x^2-p_y^2}{p_T^2}]}\rangle_{ev},\hspace{0.2cm} 
... .
\label{eq5}
\end{eqnarray}

Above deductions argued that the $v_n$ is an average of $cos(n\phi)$ first 
over particles in an event and then average over events (event-wise average) 
rather than a particle-wise average (``an average over all particles in all 
events" \cite{posk}) of $cos(n\phi)$. The particle-wise average of 
$cos(n\phi)$ can be expressed as
\begin{equation}
v_n^p=\langle\overline{cos(n\phi)}N_{evp}\rangle_{ev}/\langle N_{evp}\rangle_
{ev}
\end{equation}
(see Appendix). This is obviously different from the event-wise average. Only 
if the $\overline{cos(n\phi)}$ is independent of constrained event 
multiplicity $N_{evp}$ (i. e. constrained multiplicity plays no role in the 
average) the $v_n^p$ reduces to $v_n^e$. In fact, the $\overline{cos(n\phi)}$ 
and $N_{evp}$ correlate (even negatively correlate) with each other. This is 
because the larger event multiplicity is due to more central collision (larger 
overlap zone between colliding nuclei) and the larger overlap zone in turn 
leads to less azimuthal asymmetry. The particle-wise average does not take the 
influence of event multiplicity into account, thus it is inaccurate from 
physics point of view. Of course, for very narrow multiplicity (centrality) 
bins the particle-wise average is not very problematic. However for the wide 
multiplicity bins (such as the 0-40\% most central Au+Au collisions studied 
in \cite{phob3} and/or the minimum bias Au+Au collision data given in 
\cite{star2}) that correction is not negligible.

We have applied the parton and hadron cascade model, PACIAE \cite{sa1}, to 
calculate the charged hadron $v_2(\eta)$ in 0-40\% most central $Au+Au$ 
collisions at $\sqrt {s_{NN}}$=200 GeV by the method of event-wise and 
particle-wise averages separately. The $v_2^e(\eta)$ calculated by event-wise 
average is about 10\% larger than $v_2^p(\eta)$ calculated by particles-wise 
average. This means that $\langle\overline{cos(n\phi)}N_{evp}\rangle_{ev}$ is 
smaller than $\langle\overline{cos(n\phi)}\rangle_{ev} \langle N_{evp}\rangle
_{ev}$, i. e. there is a negative correlation between $\overline{cos(n\phi)}$ 
and $N_{evp}$. This is consistent with the physical picture given in the last 
paragraph.

In summary we have rederived the elliptic flow and other harmonics in fourier 
expansion of the particle momentum azimuthal distribution starting from the 
particle number (multiplicity) distribution in the momentum space. It is  
argued that the parameter of $n-th$ harmonic, $v_n$, should be calculated  
as an average of $con(n\phi)$ first over particles in an event and then over 
events (event-wise average). This is different from the particle-wise average 
(``an average over all particles in all events") of $cos(n\phi)$. In case of 
large centrality (multiplicity) bin the particle-wise average is not accurate  
because the influence (fluctuation) of particle multiplicity is not taken 
into account.

\begin{center} Appendix \end{center}
In this appendix the variables are assumed to be discrete for simplicity. Then 
the particle-wise average of $cos(n\phi)$ can be expressed as
\begin{equation}
v_n^p=\frac{\sum \limits_k[cos (n\phi)]_k}{\sum \limits_k 1},
\end{equation}
where $\sum \limits_k$ denotes the sum over all particles in all events. If 
the summation, both in numerator and denominator, groups according to event 
(denoted by $j$) then
\begin{equation}
v_n^p=\frac{\sum \limits_j[\sum \limits_{i\in j}[cos (n\phi)]_i]}{\sum \limits_
    j(N_{evp})_j},
\end{equation}
where $\sum \limits_j$ is the sum over events and $\sum\limits_{i\in j}$ 
denotes the sum over particles in the $j$-$th$ event. According to the 
definition of average over particle in a single event \cite{beyer} the $v_n^p$ 
can be further expressed as
\begin{equation*}
v_n^p=\frac{\sum \limits_j(\overline{cos(n\phi)})_j(N_{evp})_j}{\sum \limits_j
    (N_{evp})_j}
\end{equation*}
\begin{equation}
     =\frac{\langle\overline{cos(n\phi)}N_{evp}\rangle_{ev}}{\langle N_{evp}
\rangle_{ev}}.
\end{equation}

\begin{center} Acknowledgment \end{center}
We thank L. P. Csernai for valuable discussions. The financial supports from 
the NSFC (10475032, 10605040, and 10635020) in China and the NSFC (China)/NFR 
(Norway) collaboration are gratefully acknowledged.

\end{document}